\documentclass[journal,onecolumn]{IEEEtran}
\usepackage{color, colortbl}
\usepackage{nomencl}
\usepackage{etoolbox}
\usepackage{verbatim}
\usepackage{booktabs}
\usepackage{siunitx}
\usepackage{algorithm}
\usepackage{algpseudocode}
\usepackage{setspace}
\usepackage{wrapfig}
\usepackage{cite}
\usepackage{xcolor}
\usepackage{lipsum}
\usepackage[caption=false, font=footnotesize]{subfig}

\definecolor{orange}{rgb}{1,0.647,0}

\ifCLASSINFOpdf
  \usepackage[pdftex]{graphicx}
\else
  \usepackage[dvips]{graphicx}
\fi
\usepackage[outdir=./]{epstopdf}

\usepackage[cmex10]{amsmath}
\usepackage{amsfonts}
\usepackage{amssymb}
\makenomenclature

\begin{document}

\title{Caching Through the Skies: The Case of LEO Satellites Connected Edges for 6G and Beyond}

\author{Basem Abdellatif,
        Mostafa M. Shibl,~\IEEEmembership{Student Member, IEEE,}
        Tamer Khattab,~\IEEEmembership{Senior Member, IEEE,}
        John Tadrous,~\IEEEmembership{Senior Member, IEEE,}
        Tarek ElFouly,~\IEEEmembership{Senior Member, IEEE,}
        and Nizar Zorba,~\IEEEmembership{Senior Member, IEEE}
        
\thanks{Manuscript received Month Day, 2023; revised Month Day, 2023; accepted Month Day, 2023. (Corresponding author: Author name.)}
\thanks{Abdellatif and ElFouly are with the Department of Electrical and Computer Engineering, Tennessee Technological University, Cookeville, TN 38501, USA. (e-mails: babde006@ucr.edu; telfouly@tntech.edu).}
\thanks{Shibl, Khattab, and Zorba are with the Department of Electrical Engineering, Qatar University, Doha, Qatar. (e-mails: ma1902206@qu.edu.qa; tkhattab@ieee.org; nizarz@qu.edu.qa).}
\thanks{Tadrous is with the Department of Electrical and Computer Engineering, Gonzaga University, Spokane, WA 99202 USA (e-mail: tadrous@gonzaga.edu).}}

\markboth{IEEE Communications Magazine, Vol. VolNumber, No. Number, Month Year}
{Shibl \MakeLowercase{\textit{et al.}}: Low Earth Orbit Satellite Edge Caching for 6G and Beyond Networks}

\maketitle

\begin{abstract}
The deployment of low earth orbit (LEO) satellites with terrestrial networks can potentially increase the efficiency and reduce the cost of relaying content from a data center to a set of edge caches hosted by 6G and beyond enabled macro base stations. 
In this work, the characteristics of the communication system and the mobility of LEO satellites are thoroughly discussed to describe the channel characteristics of LEO satellites, in terms of their frequency bands, latency, Doppler shifts, fading effects, and satellite access. Three different scenarios are proposed for the relay of data from data centers to edge caches via LEO satellites, which are the ``Immediate Forward'', ``Relay and Forward'', and ``Store and Forward'' scenarios. A comparative problem formulation is utilized to obtain numerical results from simulations to demonstrate the effectiveness and validity as well as the trade-offs of the proposed system model. The simulation results indicate that the integration of LEO satellites in edge caching for 6G and beyond networks decreased the required transmission power for relaying the data from the data center to the edge caches. Future research directions based on the proposed model are discussed.
\end{abstract}


\ifCLASSOPTIONpeerreview
\begin{center} \bfseries EDICS Category: 3-BBND \end{center}
\fi
\IEEEpeerreviewmaketitle

\section{Introduction}
\label{sec:intro}
Satellite communication has become increasingly popular due to advantages such as coverage to under or uncovered areas, communication during disaster response, and technological advancements in satellite technologies \cite{ff, mc}. Moreover, low earth orbit (LEO) satellites are increasingly popular due to lower latency, cost, and higher coverage \cite{td}. Commercial LEO satellite deployments include OneWeb, Starlink, Iridium, and Telesat, with OneWeb and Starlink having large constellations of 648 and 3300 active satellites, respectively \cite{sl, ow}.

LEO and geosynchronous equatorial orbit (GEO) satellites are commercially active satellites with different characteristics in altitude, beam footprint, and orbit. LEO satellites have altitudes ranging from 300 to 1500 km and a beam footprint up to 1000 km. GEO satellites have an altitude of approximately 35,780 km and a beam footprint up to 3500 km \cite{ff}. Thus, LEO satellites are more suitable for hybridization with terrestrial networks for better cost-effectiveness and lower content placement \cite{aad}.

Terrestrial 6G networks are expected to be fast and reliable due to their high speeds, low latency, and high capacities, utilizing the THz spectrum. With data rates ranging from 100 to 200 Gbps depending on frequency and modulation technique, antenna directivity is critical for signal arrival. It is also noteworthy to mention that channel estimation relies heavily on link distance, Doppler sampling rate, and multipath resolution \cite{ds}.

By 2024, the number of internet of things devices is predicted to reach 83 billion \cite{mc}. To handle this increase, telecommunication operators are seeking solutions such as data caching at the edge and predicting end-users' popular contents, storing them in advance in 6G-and-beyond edge networks during off-peak hours. Hybrid 6G-and-beyond terrestrial and satellite communication networks have been studied to provide more flexibility and exploit the benefits of both networks \cite{aad}.

One of the main challenges for hybrid LEO-terrestrial networks is continuous availability from LEO satellites. Mobility of LEO satellites requires inter-satellite connections for content delivery, due to the high relative speed of LEO satellites. Thus, constant handovers between multiple LEO satellites might be required.

In this article, a content delivery scenario using a data center and a constellation of LEO satellites is proposed to relay content updates to a cluster of $N$ edge caches hosted by 6G-and-beyond-enabled macro base stations. The system aims to reduce the need for geographic replica hosting of data centers by using LEO satellite-enabled proactive content caching.

In the upcoming sections, Section \ref{sec:leolink} introduces the prelimiary information associated with the communication of LEO satellites. Next, Section \ref{sec:sysmod} describes the LEO satellite-based caching architecture. After that, Section \ref{sec:probform} and \ref{sec:scenarios} introduce the system model and the case study scenarios, respectively. Section \ref{sec:results} presents and discusses the case study results and tradeoffs. Consequently, Section \ref{sec:openChallenges} discusses the future research directions for the proposed LEO-based caching architecture. Finally, Section \ref{sec:conc} concludes the paper.

\section{LEO Satellite Communication}
\label{sec:leolink}
LEO satellites' low altitudes reduce propagation delays and transmission power, making them suitable for communication applications. Achieving compatibility, seamless switching, and optimal resource allocation between LEO satellite and other networks is crucial. Communication systems using LEO satellites face challenges such as limited coverage, power budget, link variation, and Doppler effect caused by high relative mobility between satellites and ground stations.

LEO satellite's operating frequency varies depending on application and spectrum allocation, including UHF, VHF, L, S, C, X, Ku, Ka, and V bands \cite{td, ys}. They typically use frequency division duplexing for higher spectral efficiency \cite{td}, but novel access schemes like frequency-hopping duplexing are also proposed to address uneven traffic demand and ensure quality of service \cite{jt}.

LEO satellites have limited power capabilities, making high efficiency crucial for communication \cite{ys}. The cost of transmission and reception depends on the power budget available for uplink and downlink to deliver content of size $B$ within a bounded delivery time $T$. Maximizing data transfer for a certain transmission power and service availability in critical fading scenarios are key design considerations \cite{td}. Also, massive MIMO transmission techniques have shown to increase data rates in recent years \cite{ly}.

As a result of the low altitude of LEO satellites ranging from 500 to 2000 km, their latency is significantly lower compared to GEO and MEO satellites. However, their high relative speed (up to 10 km/s) means a user is only served by a single satellite for a few minutes, requiring frequent handovers \cite{ys}. Thus, quick execution of handover operations is crucial to maintain quality of service, prevent data loss, and ensure efficient resource exploitation \cite{td}.

Integrating LEO satellite communication systems with terrestrial networks requires lossless handover operations and service continuity. Varying propagation delay characteristics between the networks must also be considered, as satellite networks have greater propagation delays \cite{td}.

LEO satellite's high mobility results in significantly higher Doppler shifts and frequency rate changes, which must be compensated for in communication with ground stations \cite{ys}. Various techniques, including probabilistic methods \cite{tak}, trajectory approximation \cite{ia}, and Maximum A Posteriori Doppler estimators \cite{jl}, are used to calculate the Doppler shift in LEO satellite communications.

Moreover, LEO satellite communication systems experience phase shifts caused by various factors, including phase noise, inadequate frequency synchronization, and Doppler-induced frequency changes. Techniques such as phase tracking reference signals and primary/secondary synchronization signals are used to compensate for these delays \cite{td}.

Small scale fading has little effect on satellite transmissions due to the strong line-of-sight connectivity, high signal powers, and high antenna gains. Nevertheless, large scale fading cause random variations in link quality, and modeling the fading effect depends on various factors, including LEO satellite altitude, elevation angle, ground station mobility, and weather conditions. A simple model assumes that the line-of-sight component follows a log-normal distribution and the multipath components follow a Rayleigh distribution \cite{wl}.

LEO satellites experience less atmospheric attenuation compared to GEO satellites, but it can still affect communication links between LEO satellites and ground stations. The amount of attenuation is influenced by weather conditions, operating frequency, and altitude.

Table I summarizes LEO satellite link characteristics.

\begin{table*}
\centering
\caption{\textbf{Summary of LEO Satellite Link Characteristics}}
\begin{tabular}{|c|c|c|c|}
\hline
Link Characteristic & Description & Typical Values & References \\
\hline
Frequency & Operating frequency for the uplink and & [1, 17] and [27, 75] GHz & \cite{ys, td, jt} \\
 & downlink connection. & & \\
\hline
Power Budget & Power available to provide connectivity. & [10, 50] Wh & \cite{ys, td, ly} \\
\hline
Altitude & Distance from the surface of the earth. & [500, 2000] km & \cite{ys, td} \\
\hline
Doppler Shift & Inherent changes in the frequency of the & [10, 100] kHz & \cite{ys, tak, ia, jl} \\
 & signals due to mobility. &  & \\
\hline
Phase Shift & Inherent time delays in the signal due & [$10^o$, $25^o$] & \cite{td} \\
 & to phenomena such as phase noise. & & \\
\hline
Fading & Sudden changes in signal power due to & $\sigma^2$ = [5, 20] dB & \cite{wl} \\
 & weather conditions and shadowing effects. &  & \\
\hline
Atmospheric Losses & Attenuation in the signal power due & [5, 40] dB & \cite{ck} \\
 & to gaseous particles. &  & \\
\hline
\end{tabular}
\end{table*}

\section{Proposed LEO Based Caching Architecture}
\label{sec:sysmod}
Three scenarios are envisioned for the edge caching of data for 6G-and-beyond networks using LEO satellites, which are the ``Immediate Forward'', ``Relay and Forward'', and ``Store and Forward'' scenarios. It is assumed that the data center aims to deliver $B$ units of data to the edge cache within $T$ time slots, and the satellite can simultaneously transmit, receive, and store up to $B_s$ units in its internal memory.

Additionally, the system factors in high power budget and propagation delays for the uplink and downlink, limited onboard storage, and uncertain link quality due to weather conditions and limited satellite coverage.

\textbf{Immediate Forward:} depicted in Fig. 1a. A single satellite communicates with the data center and the edge caching macro base station without using its onboard memory. The satellite relays the data to the edge cache immediately after receiving it from the data center.

\begin{figure*} 
\label{fig:scenarios}
    \centering
  \subfloat[]{%
       \includegraphics[width=172px]{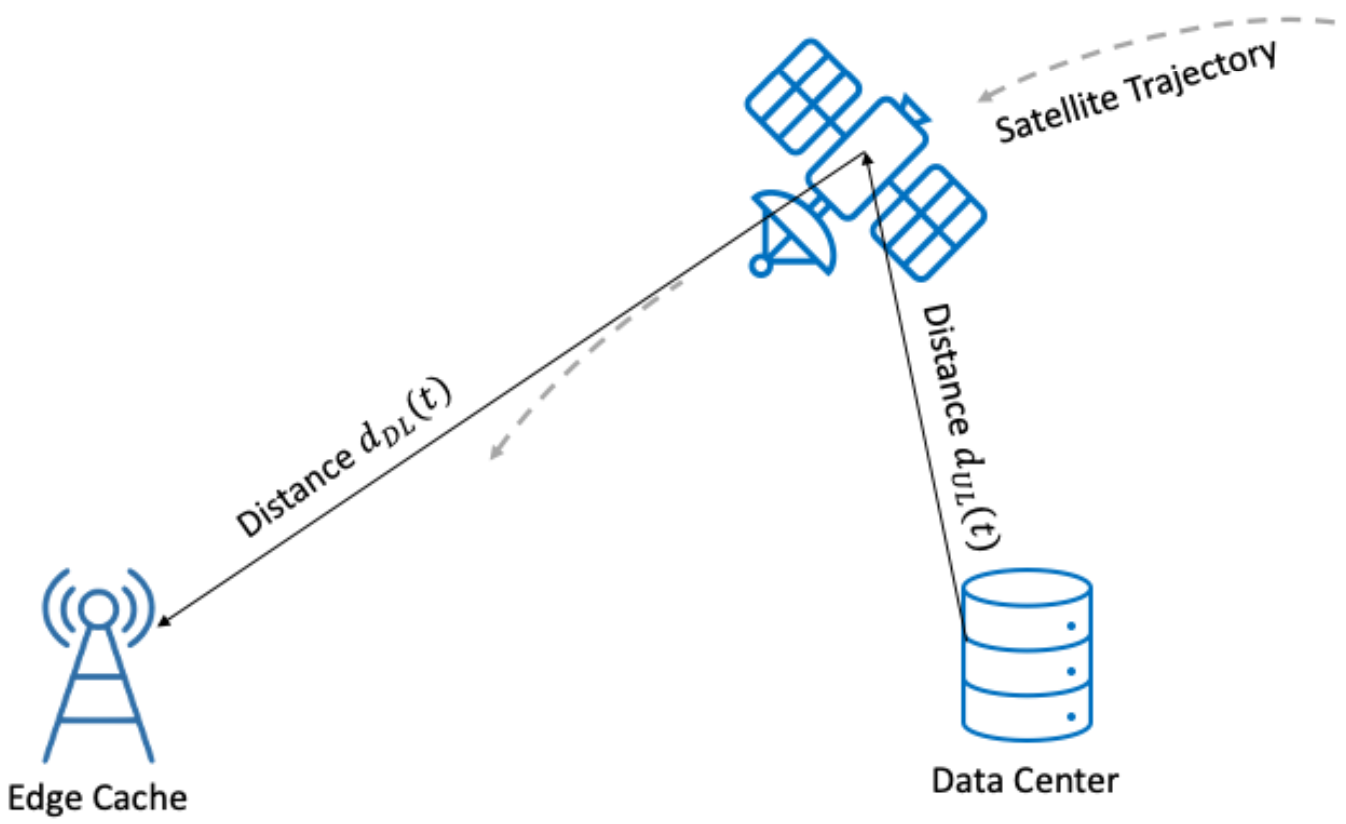}}
  \subfloat[]{%
        \includegraphics[width=165px]{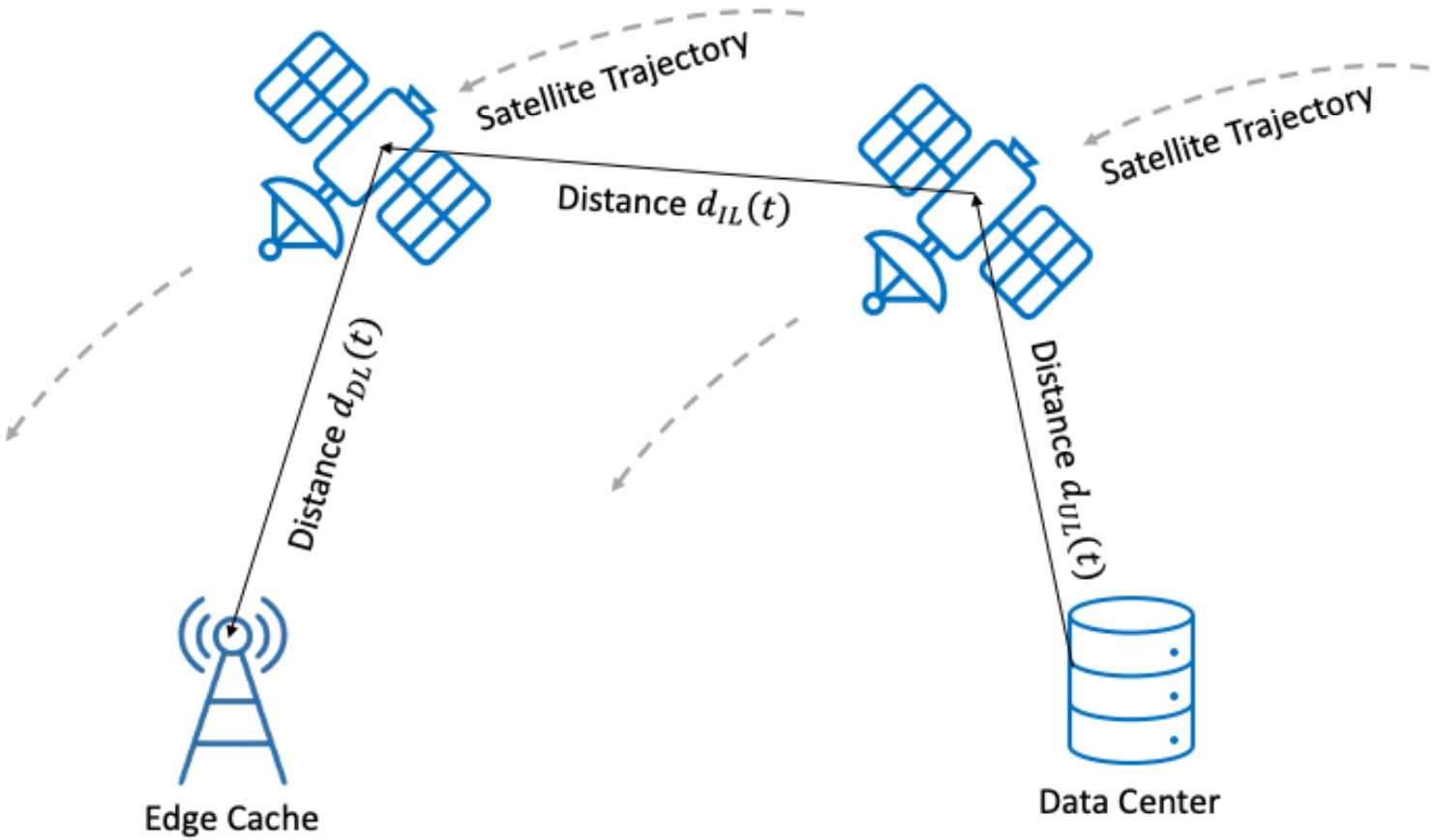}}
  \subfloat[]{%
        \includegraphics[width=174px]{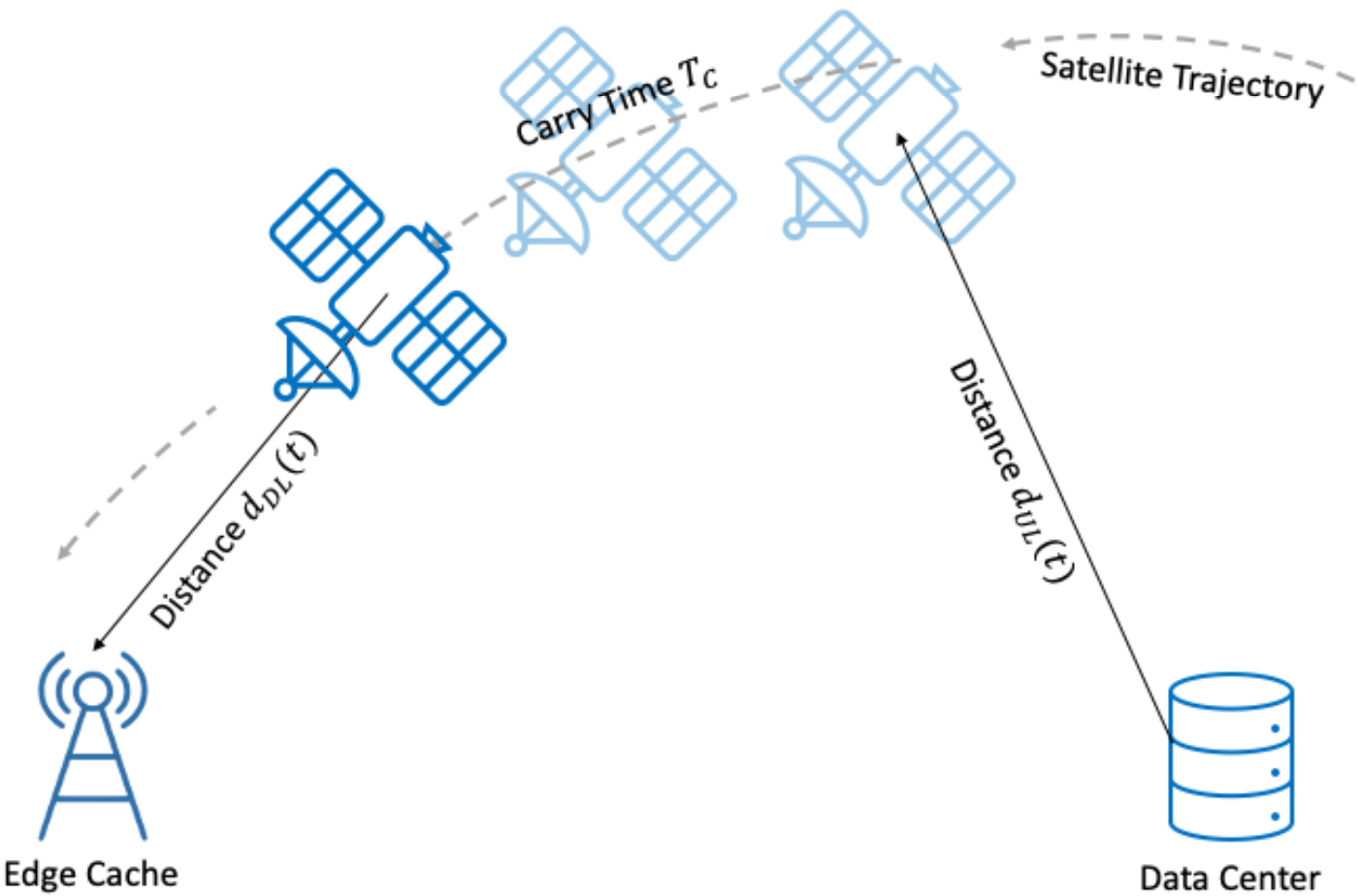}}
  \caption{(a) ``Immediate Forward'' scenario (b) ``Relay and Forward'' scenario (c) ``Store and Forward'' scenario}
\end{figure*}

\textbf{Relay and Forward:} demonstrated in Fig. 1b. Multiple satellites work together to relay data from the data center to the edge caching macro base stations when they are geographically far apart. The satellites relay data via inter-satellite links as soon as the first satellite receives the transmission from the data center. Similar to the ``Immediate Forward'' scenario, no onboard storage is required as long as the constellation provides full coverage between any two points on earth at any time.

\textbf{Store and Forward:} shown in Fig. 1c. A single satellite receives data from the data center and stores it onboard until it is in communication range with the caching edge node, where it then forwards the data. This setup incurs longer delays compared to the other scenarios, but provides a viable solution for moderate distances between the content server and the caching edge nodes, and enables the use of smaller LEO constellations that cannot guarantee full simultaneous geographic coverage.

Hybrid ``Relay and Forward'' with ``Store and Forward'' can be used depending on the size and dispersion of the constellation. In this case, relaying and storage are combined, and a utility function can be used to combine different costs of uplink, downlink, inter-satellite links power budgets, and onboard memory storage usage.

\section{System Modeling}
\label{sec:probform}
The LEO satellite closest to the data center is named the \emph{``data center satellite''}. This is the satellite that starts receiving data from the data center to be delivered to the edge cache cluster. The LEO satellite closest to the edge cache cluster is named the \emph{``edge cache satellite''}. It should be noted that the \emph{``data center satellite''} is the \emph{``edge cache satellite''} for scenarios 1 and 3 since the communication assumes no intermediate relay satellites.

$t = 1$ is the time slot when the data center can transmit to the \emph{``data center satellite''}. Thus, the time frame $1, \cdots, T$ represents the period of interest to deliver the content.  The following is a description of the key components incorporated in a mathematical system model of the proposed architectures.

\textbf{Distance model:} The distance between the data center and the \emph{``data center satellite''}, and the distance between the edge cache and the \emph{``edge cache satellite''} is varying due to the motion of the satellite. The two distances are denoted by $d_{UL}(t)$, $d_{DL}(t)$, respectively, with $t$ capturing the time dependence. These distances vary according to deterministic equations in $t$, as the satellite follows an almost deterministic orbit around Earth. Further, the distance from the data center to the edge cache cluster is denoted by $d_{C}$.

\textbf{Communication model:} As a result of the broadcast nature of wireless channels, the \emph{``edge cache satellite''} can communicate data to the entire cluster of $N$ edge caches using a single multicast transmission instead of $N$ individual unicast sessions. Accordingly, the \emph{``edge cache satellite''} consumes the same amount of power to transmit content to all edge caches as if it is serving a single edge cache. On the contrary, the data center communicates with each edge cache through a dedicated unicast TCP connection, requiring $N$ transmissions to deliver the content to the entire edge cache cluster. Long distance communication incurs propagation delays due to finite wave speeds in different media denoted by $s_C$ and $s_S$ for the medium between the data center and edge cache cluster, and free space, respectively.

\textbf{Relay LEO satellite:} In case of hauling over a long path using the LEO constellation, multiple LEO relays are utilized as introduced in scenario 2. The density of relay LEO satellites is considered to be $\lambda$ LEO satellites per meter. Consequently, the total number of LEO relays needed to haul content over a ground distance of $d_C$ is $L = \lfloor \lambda d_C \rfloor$. This means that the number of relay LEO satellites scales linearly with the communication distance.

\textbf{Power consumption and associated service rate:} In this problem, the system's total power consumption to communicate $B$ units of data to the cluster of $N$ caches in $T$ time slots is the primary performance metric. Uplink transmit power from the data center to the \emph{``data center satellite''} at time $t$ is denoted as $p_{UL}(t)$, while downlink transmit power from the \emph{``edge cache satellite''} to the edge cache cluster is denoted as $p_{DL}(t)$. Transmit power from the data center to an individual edge cache base station at time $t$ is denoted as $p_C(t)$.

Since the communication schemes between relay satellites are mainly proprietary, a formula for the power consumption at a relay satellite is unavailable. Alternatively, a power cost of $\pi$ is incurred for every relay satellite added between the data center and the \emph{``edge cache satellite''}.

\textbf{Deadline adjustment:} The service deadline $T$ has to be adjusted to account for potential propagation delays. In particular, if the communication on a path from the data center to the edge cache exhibits a propagation delay $T_{prop}$, then the service deadline has to be adjusted to $T-T_{prop}$.

\textbf{Optimization objective:} The main objective in the proposed LEO caching problem is overall power minimization, with an optimization problem formulated for three architecture scenarios and a baseline scenario, where the data center communicates with the edge cache cluster without any LEO satellites. The power cost metric prioritizes the limited resources of LEO satellites, utilizing a weighting factor $\alpha \in[0,1]$ to emphasize their power cost and assign $1-\alpha$ to the data center's power cost.

\section{Case Study Scenarios}
\label{sec:scenarios}
To compare the baseline and proposed scenarios, a basic technique can be used to uniformly distribute data over the available time period. For example, with $B = 400$ data units and $T_0 = 200$ time slots, the technique would transmit $2$ data units per time slot at a fixed rate.

The LEO satellite to data center and LEO satellite to edge cache channels assume a Rician distribution, resulting in a non-central Chi-squared distribution for power gain.

The simulations plot required signal-to-noise ratio (SNR) against percentage of data sent via the LEO constellation, with fixed unity noise power. Varying chi-squared parameters capture different channel conditions. Each scenario indicates an optimal amount of data to send via LEO satellites depending on channel conditions. These simulations aim to demonstrate the caching advantage of LEO satellites. As a result, derivations are omitted, and some values are assumed based on existing systems and literature.

\subsection{Baseline Cost Minimization}
The data center supplies content updates of $B$ data units to each of the $N$ edge caches individually using unicast transmission. There is no LEO satellite support to communicate the data. Therefore, the propagation delay to a single 6G-and-beyond base station is $d_C/s_C$. This means that the data center has to deliver $B$ units of data to an edge cache within $T_0 = \lfloor T/N - d_C/s_C \rfloor$.

\subsection{Scenario 1: Immediate Forward Model}
In this scenario, the single LEO satellite can help deliver $B_s \leq B$ of the data to the entire edge cache cluster through means of broadcasting. Data delivery through the LEO satellite incurs an average propagation delay that is the sum of the mean of the uplink and downlink delays. Accordingly, the communication deadline is adjusted to account for such propagation delays. Thus, the total minimum power cost for scenario 1 is computed based on the weighting factors ($\alpha$ and $1 - \alpha$), and the minimum operating power of the LEO satellite and the data center.

\subsection{Scenario 2: Relay and Forward Model}
When the data center and 6G-and-beyond edge cache cluster are located within a distance larger than the visibility area of a single LEO satellite, a constellation of LEO satellites relay this data from the \emph{``data center satellite''} to the \emph{``edge cache satellite''}.

In this scenario, the total propagation delay is decomposed into the following components:

\begin{itemize}
    \item from the data center to the \emph{``data center satellite''}.
    \item through the relays.
    \item from the \emph{``edge cache satellite''} to the edge caches.
    \item from the data center to the edge cache cluster.
\end{itemize}

It is noteworthy to mention that the relay constellation employs ``Store and Forward'' in the sense that the data center first uploads $B_s$ data units to the \emph{``data center satellite''}, followed by the relay of data through the constellation. Consequently, the $T$-slot communication frame is divided into the following segments: from $1$ to $T_{2UL}$ for the data center to upload data, from $T_{2UL}$ to $T_{2R}$ for the data to be relayed to the \emph{``edge cache satellite''}, and from $T_{2R}$ to $T_{2DL}$ for the \emph{``edge cache satellite''} to broadcast the $B_s$ data units to the edge cache cluster.

In this scenario, the minimum operating power will be found based on the weighting factors and the minimum operating power of the LEO satellite and the data center, and the number of relay satellites, the relay density per unit distance, and the LEO-to-LEO power cost.

\subsection{Scenario 3: Store and Forward Model}
In this scenario, a single LEO satellite receives $B_s$ data units from the data center while it is in the visibility region. After traveling a distance proportional to $d_C$ (the ground distance between the data center and the edge caches) with a velocity $v$, the LEO satellite then broadcasts the $B_s$ data units to the $N$ edge caches.

To meet the $T$-slot service window while accounting for propagation delays together with the LEO satellite travel time, the service time frame is divided as follows: from $t=1$ to $T_{3UL}$ for the data center to upload $B_s$ data units to the LEO satellite, from $T_{3UL}$ to $T_{3V}$ for the LEO satellite to travel to the edge cache cluster, and from $T_{3V}$ to $T_{3DL}$ for the LEO satellite to broadcast the data to the 6G-and-beyond base stations. Thus, the incurred propagation delays in this scenario decompose to the following:

\begin{itemize}
    \item from data center to the LEO satellite.
    \item from the LEO satellite to 6G-and-beyond base stations.
    \item from the data center to the base stations.
\end{itemize}

Scenario 3's power cost is calculated using the weighting factors and the minimum operating power of LEO satellite and data center. The cost function also considers storage costs in satellite caching memory to prevent long-term memory occupancy. The ``Relay and Forward'' scenario may be more cost-efficient than the ``Store and Forward'' based on delay and storage balance, depending on the constellation size and satellite/data center location.

\section{Results and Tradeoffs}
\label{sec:results}
In the simulation results, the LEO satellite is assumed to have an altitude of $1200 \: km$, and an edge cache cluster of only $2$ edge caches is located at a distance of $60 \: km$ from the data center. Also, the signals are assumed to be traveling at the speed of light. Fig. 2 shows the optimal amount of data to be relayed through LEO satellites from a power saving point of view for all scenarios, and compares the results between the baseline and the different scenarios.

In all scenarios, $B = 400$ chunks of data, where a chunk is of size $1400$ bytes (typical size of a TCP packet), are to be allocated in a total time of $T = 200 \: ms$. In the second scenario, it is assumed that the size of the LEO constellation is 5 satellites. In the third scenario, it is assumed that a single LEO satellite that stores the data and travels to the edge cache at a ground speed $v = 10 \: km/s$.

\begin{figure*} 
\label{fig:results}
    \centering
  \subfloat[]{%
       \includegraphics[width=172px]{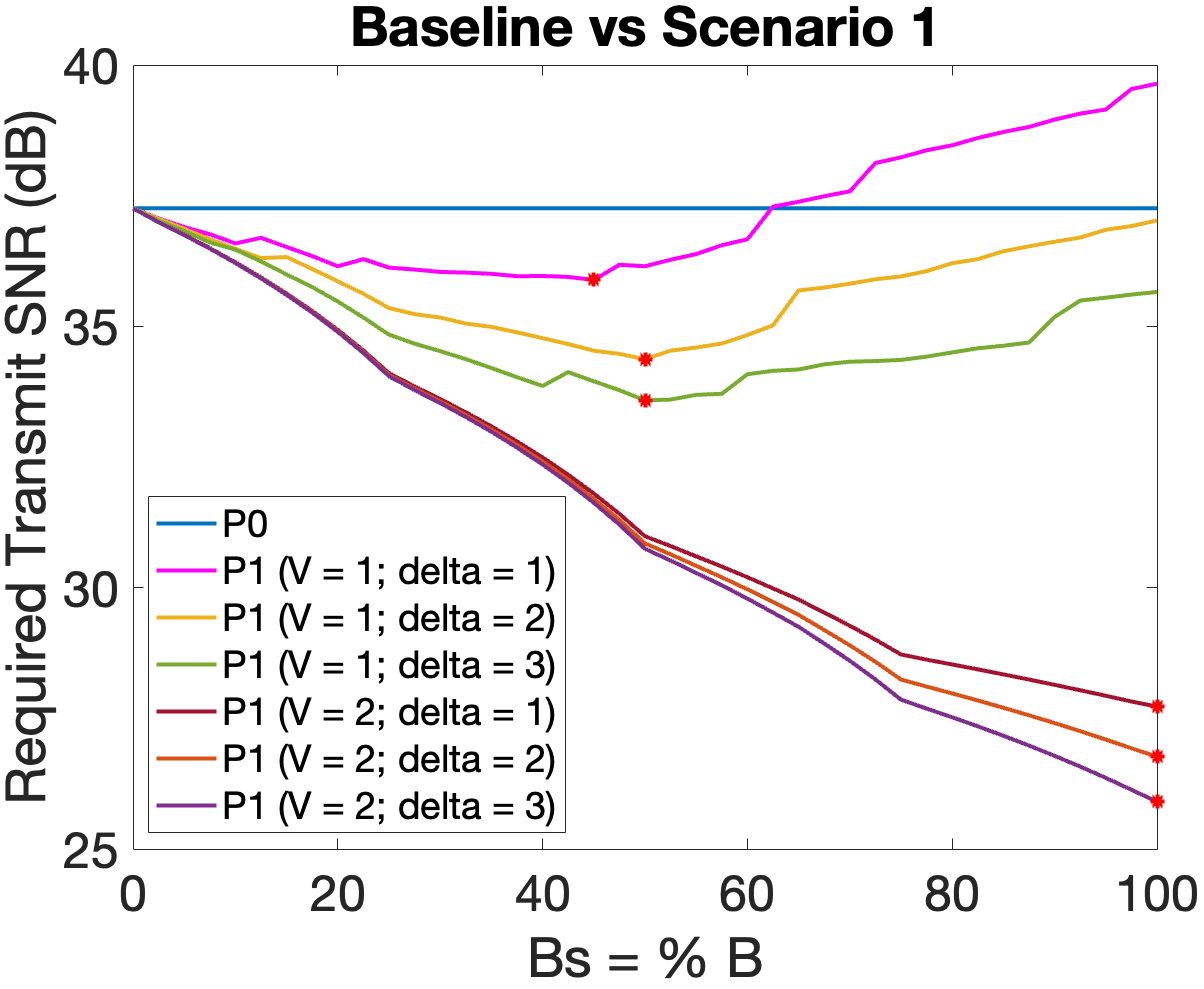}}
  \subfloat[]{%
        \includegraphics[width=165px]{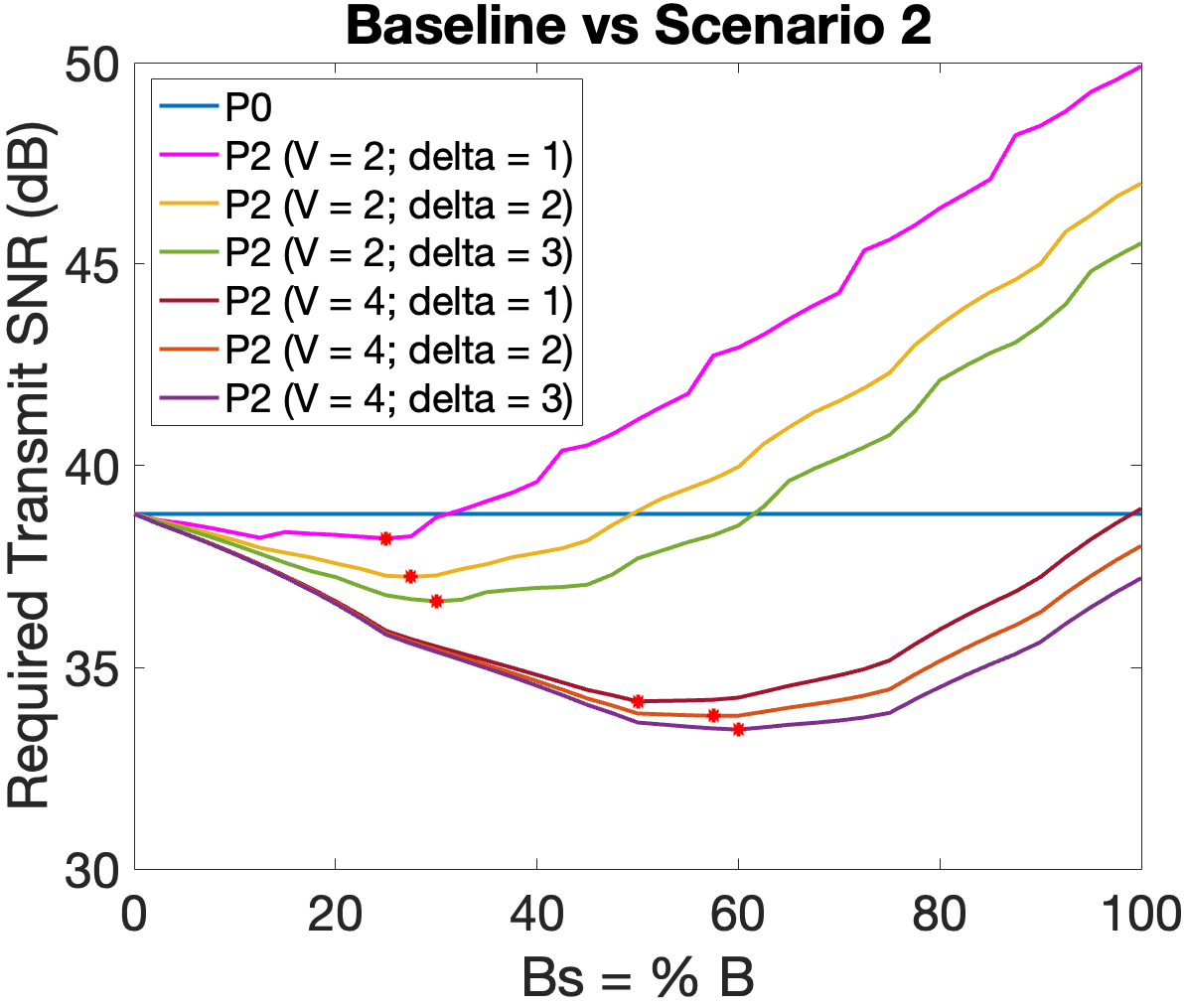}}
  \subfloat[]{%
        \includegraphics[width=174px]{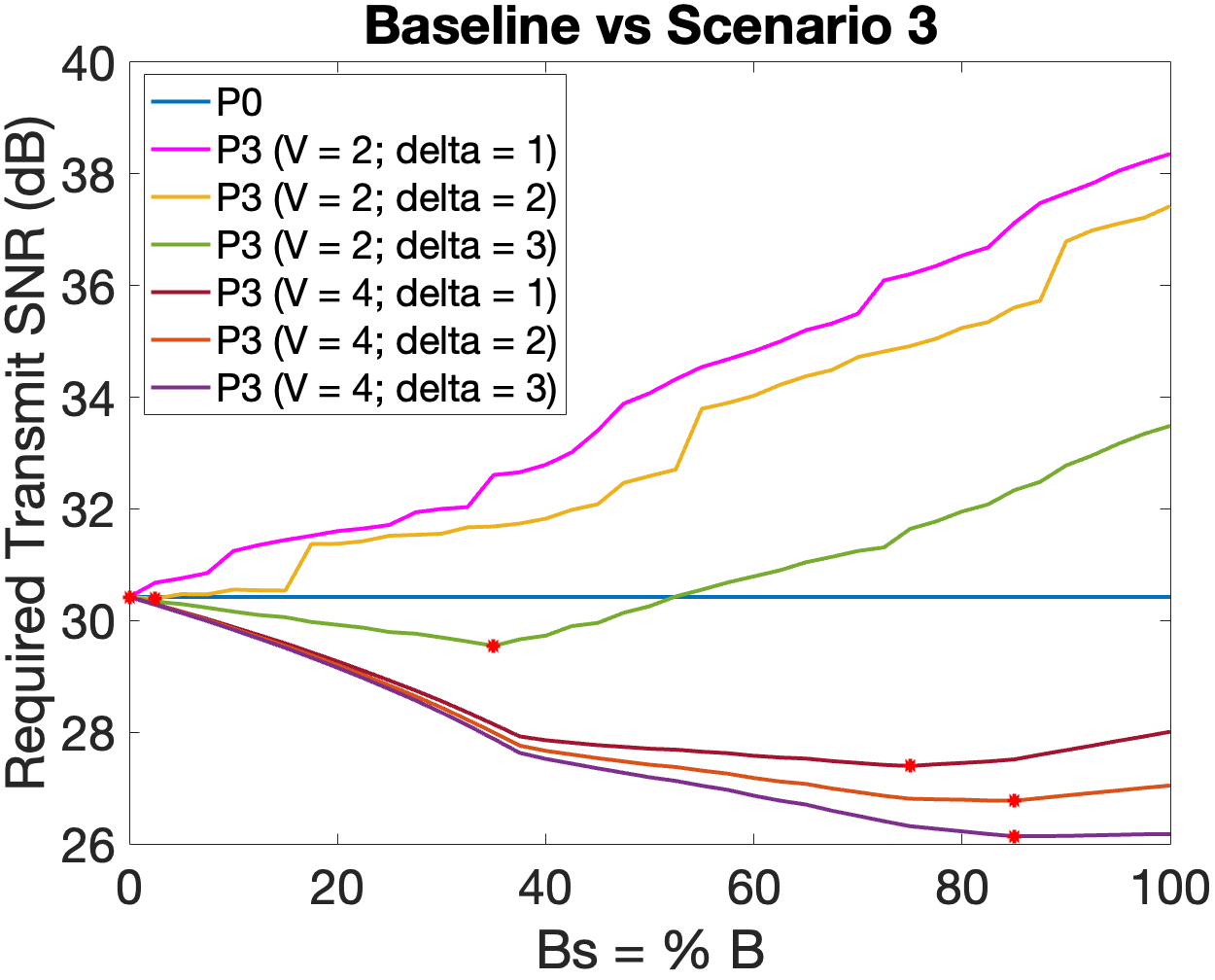}}
  \caption{Comparison between the baseline and (a) scenario 1 (b) scenario 2 (c) scenario 3}
\end{figure*}

In the ``Immediate Forward'' scenario, there is always an optimal amount of data to be relayed through the LEO satellite. While in the ``Relay and Forward'' and the ``Store and Forward'' scenarios, the baseline would give better results in the presence of weak channel characteristics.

In addition, it can be seen that the required SNR values in the ``Relay and Forward'' scenario are larger than those in the ``Immediate Forward'' scenario due to requiring multiple LEO satellites to relay the data, which utilizes more power. In the ``Store and Forward'' scenario, as the travel time of the LEO satellite from the data center side to the edge cache side increases, the data center has less time to upload the data. In order to upload the data in a short time period, the total power consumption increases.

Thus, it can be deduced that there is a tradeoff between the optimal amount of data to be relayed through LEO satellites and the channel characteristics. As the channel characteristics between the data center and the LEO satellite improves, the amount of data to be relayed through the satellite increases, and then required transmit SNR decreases. In addition, in some scenarios, the baseline might provide better results if the channel characteristics becomes increasingly weak.

Moreover, it is worth noting that there are other factors that would affect the power consumption excluding the channel characteristics and the distance between the data center and the edge caches, such as the number of edge caches, the number of LEO satellites in the ``Relay and Forward'' scenario, the speed of the LEO satellite in the ``Store and Forward'' scenario, and the amount of data to be delivered.

\section{Open Challenges}
\label{sec:openChallenges}
The incorporation of LEO satellite communications with terrestrial networks is a promising research area due to the possible benefits of a hybrid model between satellite and terrestrial networks. A number of challenges that might haul the advancement of such integration provide a number of future research directions.

Firstly, in all LEO satellite relay scenarios, it was assumed that data comes from a single data center. In real scenarios, multiple data centers with different delivery requirements may communicate with the LEO satellite simultaneously. Thus, when the data centers have different delivery deadlines, an efficient scheduling algorithm is required to meet the different deadlines.

Another challenge is the limited amount of onboard storage in LEO satellites. There could possibly be a situation where a LEO satellite is in communication with multiple data centers. Thus, the LEO satellite has to efficiently manage and optimize the usage of its storage to maximize its benefits.

Furthermore, challenges can also arise from the point of view of the data center. For instance, the data center must determine the optimal data amount for satellite relay. In the above formulation, a simple cost function that only incorporates the transmission power was utilized. On the contrary, in more realistic scenarios, a more complex cost function should be considered with optimization variables, such as storage cost at the LEO satellite, including storage size and storage usage duration costs.

Additionally, another challenge is the optimal transmission rate between data center and LEO satellite. In the above simulation, the rate is fixed, but in reality, scheduling algorithms could be applied to adjust the rate during good and bad channel conditions to avoid power wastage, and for higher efficiency.

Another possible research direction for the proposed architecture is the integration of artificial intelligence for cache management. Artificial intelligence algorithms can predict content popularity based on demand patterns, and pre-fetch requested data from the core network to edge caching nodes to reduce latency and congestion.

The proposed system architectures can be enhanced through dynamic load balancing algorithms, by distributing content requests efficiently among available caching nodes based on content type, user groups, or geographical regions. This can prevent overloading and optimize resource usage.

These challenges must be addressed to fully benefit from integrating LEO satellites with terrestrial networks, creating promising research problems with many potential benefits.

\section{Conclusion}
\label{sec:conc}
In this article, the benefits that could be gained from integrating satellite communications, specifically LEO satellites, with terrestrial networks were briefly introduced. Three different scenarios were discussed, which are the ``Immediate Forward'', ``Relay and Forward'', and ``Store and Forward'' scenarios, where LEO satellites are utilized in relaying data from a data center to a number of edge caches. The three scenarios span most of the practical relaying cases, where the distance between the data center and the edge caches is the major decision factor.

Due to the broadcast nature of wireless communication channels, LEO satellites do not have to transmit the data to each edge cache separately as in the baseline case, resulting in a reduced required transmit SNR. Simple simulations were performed to show the impact of LEO satellites on the communication process between the data center and edge caches, demonstrating that caching through LEO satellites reduces the required transmit SNR compared to a baseline scenario.

Finally, some of the open challenges for integrating LEO satellite communications with terrestrial networks include relaying data from multiple data centers, limited onboard storage of LEO satellites, and calculating the optimal transmission rate. These challenges offer opportunities for further research to achieve more conclusive results on the integration of LEO satellite communications with terrestrial networks.

\bibliographystyle{IEEEtran}
\bibliography{refs}

\begin{IEEEbiography}
[{\includegraphics[width=1in,height=1.25in,clip,keepaspectratio]{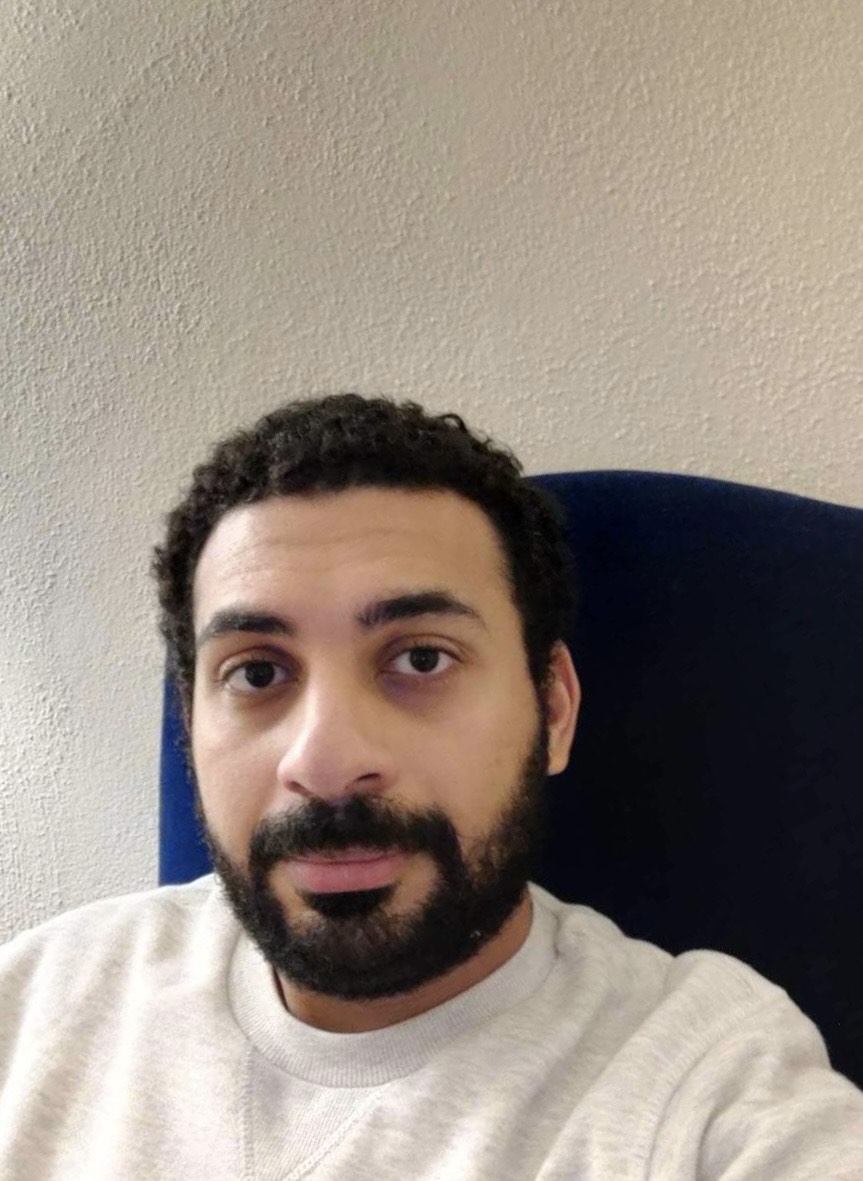}}]
{Basem Abdellatif}
received M.Sc. in Computer Science, University of California, Riverside, and M.Sc. in Electrical Engineering, Nile University, Egypt. Also, he received B.Sc. in Electrical Engineering, Cairo University, Egypt.
\end{IEEEbiography}

\begin{IEEEbiography}
[{\includegraphics[width=1in,height=1.25in,clip,keepaspectratio]{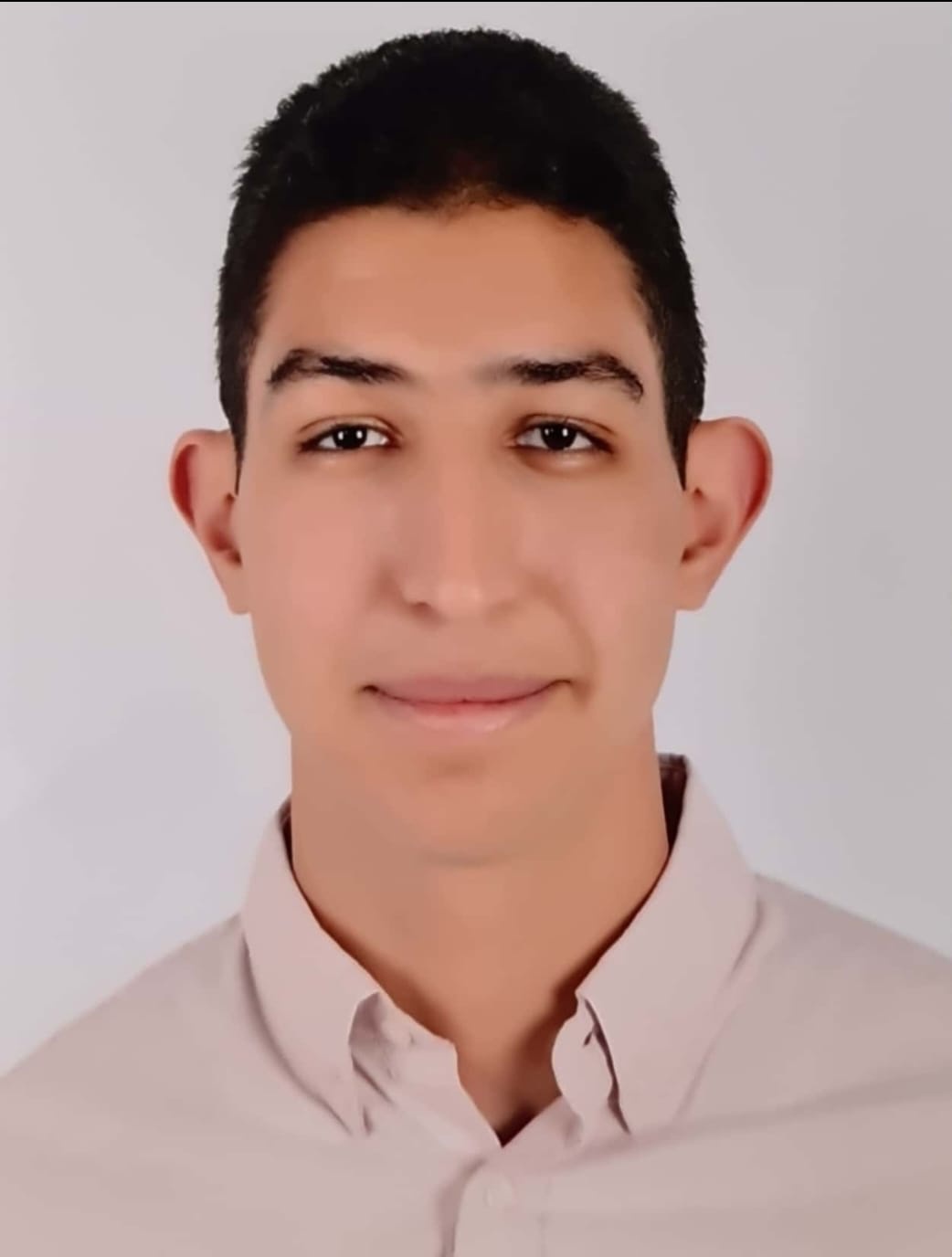}}]
{Mostafa M. Shibl}
(Student Member) is an Honors Electrical Engineering Student in Qatar University, Doha, Qatar. He is working as a research assistant on a funded project by Qatar National Research Fund (QNRF) and The Scientific and Technological Research Institution of Turkey (TÜBITAK). Currently, he has publications in the research areas of management of electric vehicles charging and battery monitoring, machine and reinforcement learning, satellite communication, and IoT. He is the recipient of the Dean’s List and Vice President’s List awards from Qatar University.
\end{IEEEbiography}

\begin{IEEEbiography}
[{\includegraphics[width=1in,height=1.25in,clip,keepaspectratio]{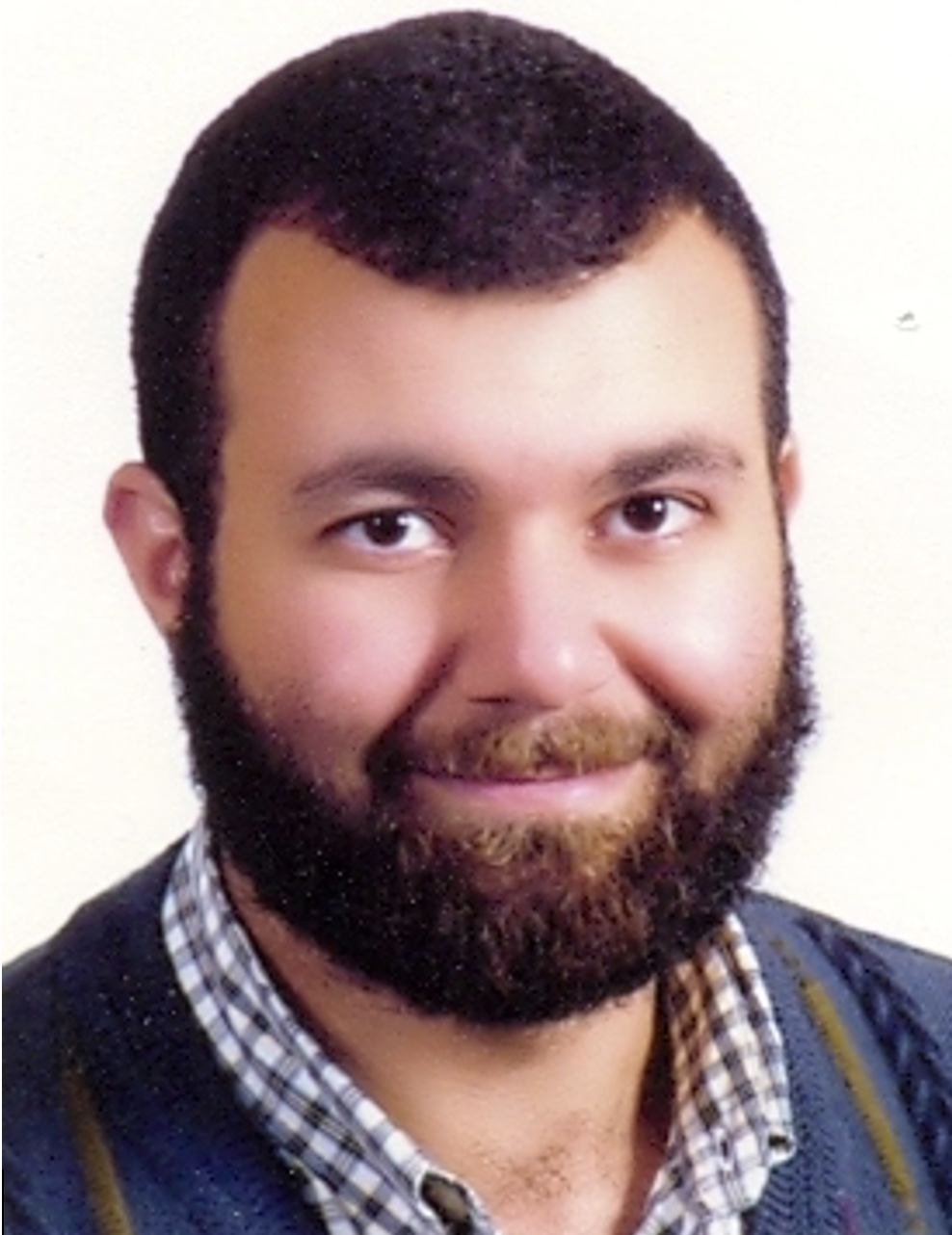}}]
{Tamer Khattab}
(M’ 94, SM’ 18) received the B.Sc. and M.Sc. degrees in electronics and communications engineering from Cairo University, Giza, Egypt, in 1993 and 1999, respectively and the Ph.D. degree in electrical and computer engineering from The University of British Columbia (UBC), Vancouver, BC, Canada, in 2007. From 1994 to 1999, he was with IBM wtc, Giza, Egypt, as a development team lead. From 2000 to 2003, he was with Nokia (formerly Alcatel Canada Inc.), Burnaby, BC, Canada, as a senior member of the technical staff. He joined Qatar University (QU) in 2007, where he is currently a Professor of Electrical Engineering and the Director of the Center for Excellence in Teaching and Learning (CETL). He is also a senior member of the technical staff with Qatar Mobility Innovation Center (QMIC).  His research interests cover satellite communications, quantum communications, information theory, RF sensing techniques and optical communications.
\end{IEEEbiography}

\begin{IEEEbiography}
[{\includegraphics[width=1in,height=1.25in,clip,keepaspectratio]{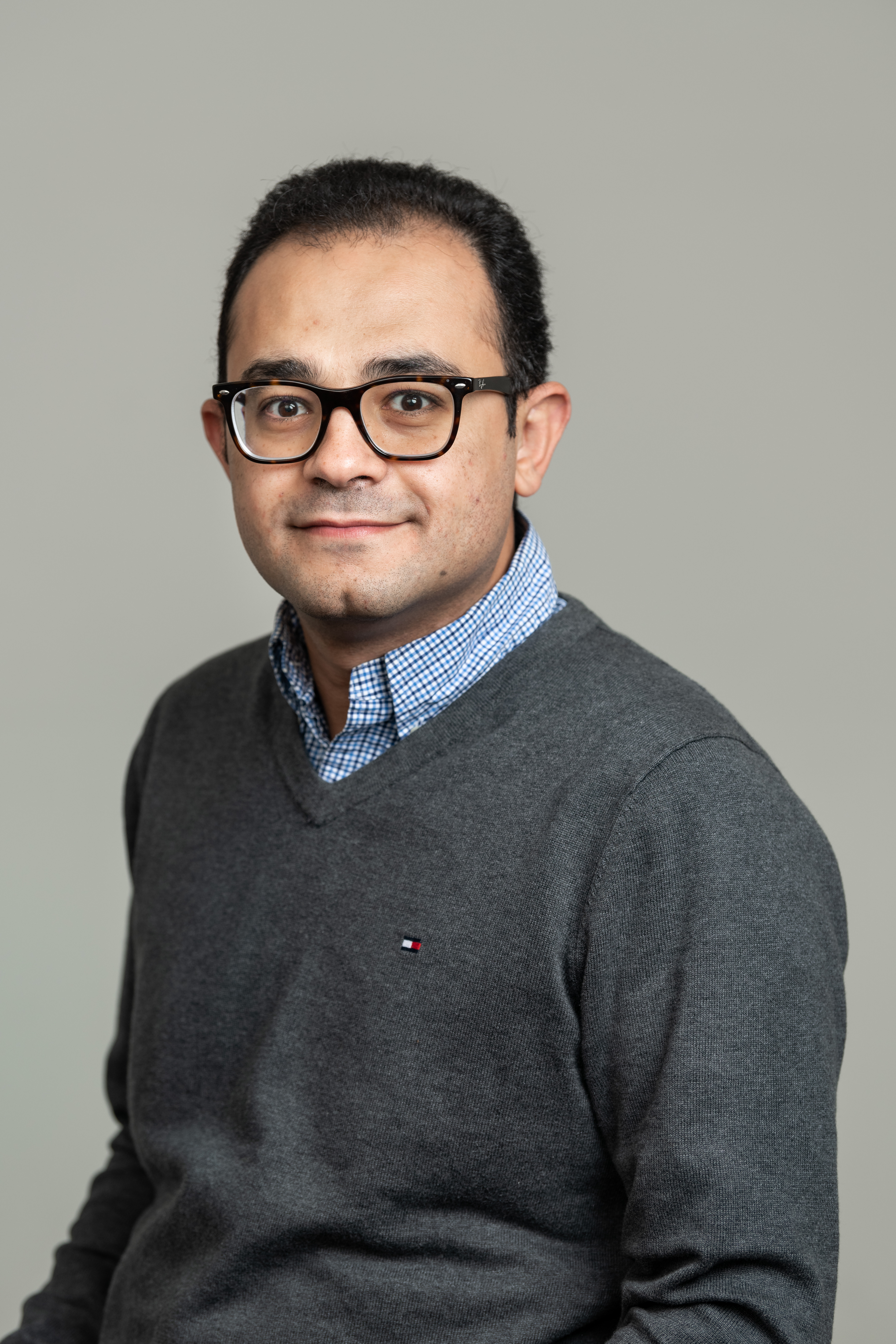}}]
{John Tadrous}
is an associate professor of electrical and computer engineering at Gonzaga University. He received a Ph.D. degree in electrical engineering from the ECE Department at The Ohio State University, an MSc degree in wireless communications from the Center of Information Technology at Nile University, and a BSc degree from the EE Department at Cairo University. He was a post-doctoral research associate with the ECE Department at Rice University between May 2014 and August 2016. In 2020, Dr. Tadrous was elevated to a Senior Member of the IEEE. In addition, he received Gonzaga University’s Faculty Award for Professional Contributions. His research interests include modeling and analysis of human behavior’s impact on data networks in various timescales from seconds to hours, and how to harness that behavior for improved network resource management. Dr. Tadrous’ served as a technical program committee member for several conferences such as Mobihoc, COMSNETS, and WiOpt.
\end{IEEEbiography}

\begin{IEEEbiography}
[{\includegraphics[width=1in,height=1.25in,clip,keepaspectratio]{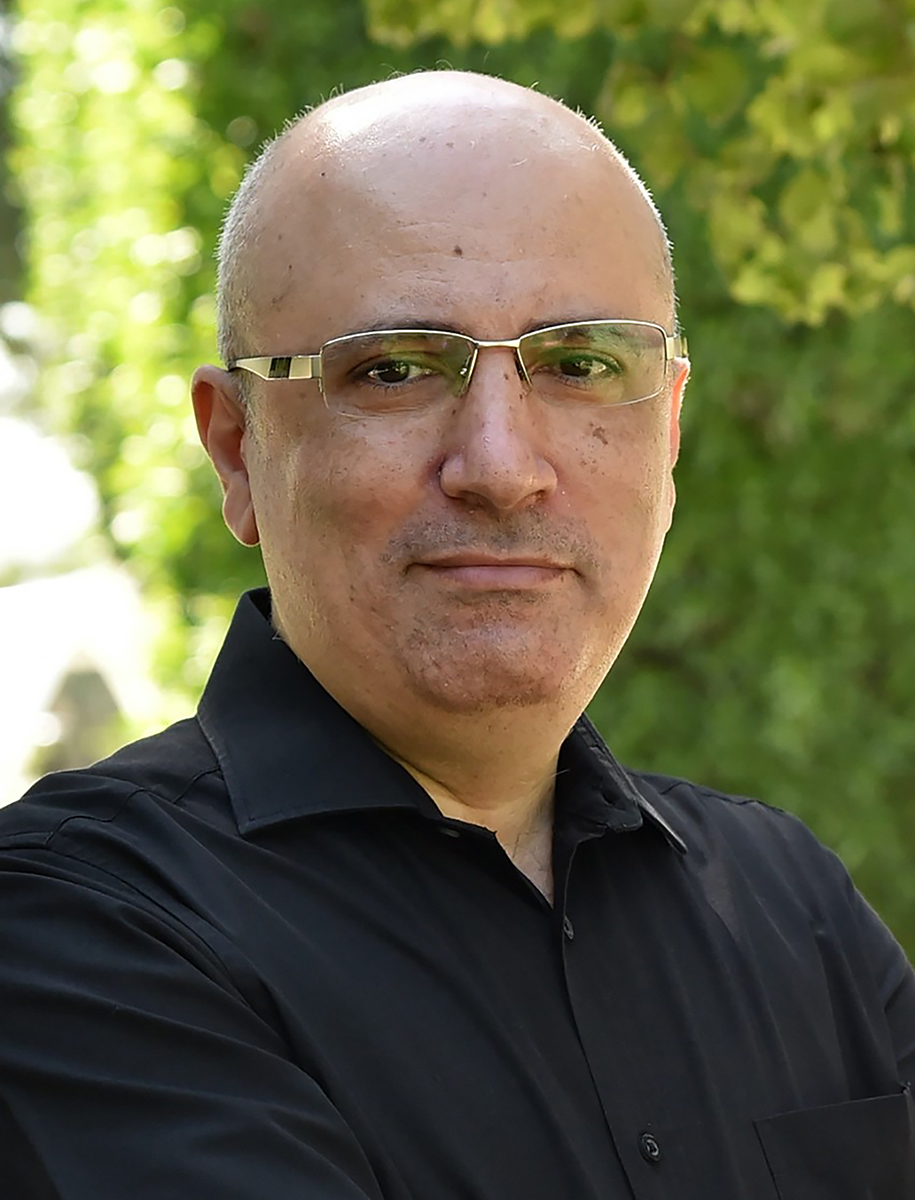}}]
{Tarek ElFouly}
(M’ 06, SM’ 13) received his DEA and PhD from the University of Franche Comte in France, in 1996 and 2000 respectively. He has worked as an assistant professor at the University of Ain Shams Cairo Egypt before joining Qatar University. He is currently an associate Professor at Tennessee Technological University, Tennessee, USA. He has over 12 years of experience in computer network research. Dr. Elfouly published over 70 papers. Dr. Elfouly supervised many post graduate students and served as an examiner for many others. Dr. Elfouly current research interests are EV battery safety, Wireless sensing networks, medical applications of machine learning. His research interests also include network security and protocols, physical layer security and wireless sensor networks especially in the field of structural health monitoring. His projects won many national and regional awards.
\end{IEEEbiography}

\begin{IEEEbiography}
[{\includegraphics[width=1in,height=1.25in,clip,keepaspectratio]{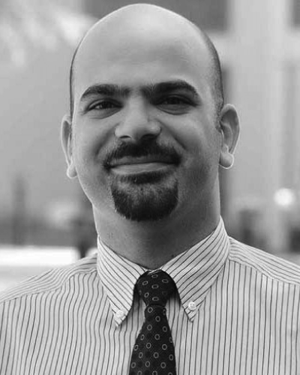}}]
{Nizar Zorba}
(Senior Member, IEEE) received the B.Sc. degree in electrical engineering from JUST University, Jordan, in 2002, and the Ph.D. degree in signal processing for communications from UPC Barcelona, Barcelona, Spain, in 2007. He is currently a Professor with the Department of Electrical Engineering, Qatar University, Doha, Qatar. He has authored five international patents and coauthored more than 150 papers in peer-reviewed journals and international conferences. Dr. Zorba is Symposium Chair of IEEE Globecom 2023, ICC 2023, Globecom 2021, and ICC 2019. He is also the Chair of the IEEE ComSoc Communication Systems Integration and Modeling Technical Committee (TC CSIM), Area Editor of the IEEE Communications Letters, Editor of IEEE IoT Magazine and Editor of IEEE TCCN.
\end{IEEEbiography}

\end{document}